\documentclass[aps,prx,twocolumn,showpacs,superscriptaddress,floatfix]{revtex4-1}  

\usepackage{ucs}
\usepackage[utf8x]{inputenc}
\usepackage{natbib}
\usepackage{graphicx}  
\usepackage{bm}        
\usepackage{amssymb}   
\usepackage{amsmath}
\usepackage{color}
\usepackage{CJK}
\usepackage{times}
\usepackage{verbatim}
\usepackage{mathrsfs}
\usepackage{hyperref}
\hypersetup{colorlinks = True, urlcolor=blue, linkcolor=blue, citecolor=blue}

\usepackage{cancel}
\begin{document}

\begin{CJK}{UTF8}{bsmi}
\title{Emergent Snake Magnetic Domains in Canted Kagome  Ice}
\author{Wen-Han Kao (高文瀚)}
\affiliation{Department of Physics and Center for Theoretical Physics, National Taiwan University, Taipei 10607, Taiwan}
\author{Gia-Wei Chern}
\affiliation{Department of Physics, University of Virginia, Charlottesville, Virginia 22904, USA}
\author{Ying-Jer~Kao (高英哲) } 
\email{yjkao@phys.ntu.edu.tw}
\affiliation{Department of Physics and Center for Theoretical Physics, National Taiwan University, Taipei 10607, Taiwan}
\affiliation{National Center for Theoretical Sciences, National Tsing Hua University, Hsinchu 30013, Taiwan}
\affiliation{Department of Physics, Boston University, 590 Commonwealth Avenue, Boston, Massachusetts 02215, USA}
\date{\today}

\begin{abstract}
We study the two-dimensional kagome-ice model derived from a pyrochlore lattice with second- and third-neighbor interactions.
The canted moments align along the local $\langle 111 \rangle$ axes of the pyrochlore and respond to both in-plane and out-of-plane external fields.
%
%
%
We find that the combination of further-neighbor interactions together with the external fields introduces a rich phase diagram with different spin textures. 
Close to the phase boundaries, metastable $\textit{"snake"}$ domains emerge with extremely long relaxation time.
Our kinetic Monte Carlo analysis of the magnetic-field quench process from saturated state shows unusually slow dynamics.
Despite that the interior spins are almost frozen in snake domains, the spins on the edge are free to fluctuate locally, leading to frequent creation and annihilation of monopole-anti-monopole bound states.
%
%
Once the domains are formed, these excitations are localized and can hardly propagate due to the energy barrier of snakes. 
The emergence of such snake domains may shed light on the  experimental observation of dipolar spin ice under tilted fields, and provide a new strategy to manipulate both spin and charge textures in artificial spin ice. 
\end{abstract}

\maketitle
\end{CJK}






\section{Introduction}\label{Introduction}
%
%
Metastability  in physical systems indicates a rough free energy landscape with multiple nearly degenerate local minima. 
Normally, it requires random disorders for the metastable behavior to occur, with spin glass as the most prominent example~\cite{Edwards_1975}. 
On the other hand, disorder-free geometrically frustrated spin-ice compounds, such as Ho$_{2}$Ti$_{2}$O$_{7}$ and Dy$_{2}$Ti$_{2}$O$_{7}$ also suffer from spin freezing  at low temperature due to the slow dynamics originated from the extensive quasi-degenerate low-energy states,  and the topological constraint known as the $\textit{ice rule}$~\cite{Harris_1997, Ramirez_1999}. 
For pyrochlore lattices with corner-sharing tetrahedra, the  ice rule dictates that the  spins  in each tetrahedron should obey the two-in-two-out constraint, resulting in a divergence-free flux field on the dual diamond lattice. 
The ice-rule constraint originated from the effective nearest-neighbor coupling of spins can be regarded as a conservation law of emergent gauge fields, giving rise to the anisotropic dipolar correlation~\cite{Isakov2004}. 
The long-range dipolar interaction in the more sophisticated dipolar spin ice (DSI) model merely provides correction to the projectively equivalent band of low-energy states~\cite{Isakov2005}; the quasi-degenerate ice manifold remains as a stable low-temperature phase over a wide temperature window~\cite{melko2001}. 

Remarkably, excitations upon this ice-rule manifold are fractionalized quasiparticles with magnetic Coulomb interaction, leading to the resemblance of magnetic monopole~\cite{castelnovo2008magnetic,Morris2009} and the emergence of unusual spin liquid called the Coulomb phase~\cite{Henley_2010, Fennell_2009}. 
Furthermore, the theoretical proposal of spin fragmentation shows that Coulomb phase can occur naturally from non-divergence free systems and coexist with long-range ordered monopoles~\cite{brooks2014magnetic}. 
Experimental signatures of spin fragmentation can be found in various systems such as quantum spin ice Nd$_{2}$Zr$_{2}$O$_{7}$~\cite{Petit_2016}, tripod kagome compound Dy$_{3}$Mg$_{2}$Sb$_{3}$O$_{14}$~\cite{dun2016magnetic,paddison2016emergent}, and thermally active artificial kagome array of GdCo alloy~\cite{canals2016fragmentation}. 
The interplay between geometrical frustration and interacting topological charges brings about complex spin textures~\cite{udagawa2016out, mizoguchi2017clustering, chioar2016ground, smerald2016topological} and intriguing out-of-equilibrium phenomena~\cite{mellado2010dynamics, chioar2014kinetic, montaigne2014}.

Under a [111] magnetic field, the pyrochlore spin ice  exhibits a two-stage magnetization process~\cite{matsuhira2002new}. 
The first plateau refers to kagome-ice phase with residual spin degeneracy, and the second plateau is a long-range ordered state which can be interpreted as a magnetic monopole condensate~\cite{castelnovo2008magnetic}. 
Neutron scattering experiments reveals the pinch point singularity of kagome-ice phase by applying magnetic field along [111] crystallographic direction on Ho$_{2}$Ti$_{2}$O$_{7}$, yet an unusual critical scattering appears when   the field is slightly tilted away from this direction~\cite{fennell2007pinch}. 
By using the dipolar spin-ice model with combination of fields, it is shown that the critical scattering is ascribable to the long-range ordered $\mathbf{q}$ = X state, and the lack of Bragg peak may be attributed to partial ordering in the real material~\cite{Kao2016}.

In this paper, we study the  interplay between further-neighbor interactions and tilted magnetic field in kagome ice in order to understand
possible mechanisms that prevent ordering.
Specifically, we study here a two-dimensional classical kagome-ice system with  magnetic moments aligning along the local $\langle 111 \rangle$ axes originated from the pyrochlore lattice.
Due to this canting, the moments can respond to both in-plane and out-of-plane  fields, allowing us to explore the rich phase diagrams by tuning both the second- and third-neighbor interactions, and the fields.
In particular, at the boundary between different phases we find metastable winding snake domains emerge with local excitations of  monopole-anti-monopole pairs.
Unlike in the conventional spin ice, these excitations are immobilized and can not proliferate to destabilize the snake domains. 
This leads to observable signatures in the neutron experiments and should provide a new perspective on the interpretation of  experimental data.

The rest of the paper is organized as follows: In Sec.~\ref{The Model} we introduce the model Hamiltonian and the geometry of canted moment and external fields. 
In Sec.~\ref{Ground States} we present the equilibrium Monte Carlo results and the phase diagrams of various ground states. 
To show the relaxation process from saturated state and the emergence of snake domains, we perform the rejection-free kinetic Monte Carlo simulation. 
The method and results are described in Sec.~\ref{Relaxation process}. 
Then, in Sec.~\ref{The snake domains}, we discuss the energetics of snake domain and the localized excitation on its periphery, along with possible experimental signatures in spin-ice crystals or artificial spin ices. 
We end with concluding remarks in the last section.

\begin{figure}
     \includegraphics[width=0.9\columnwidth]{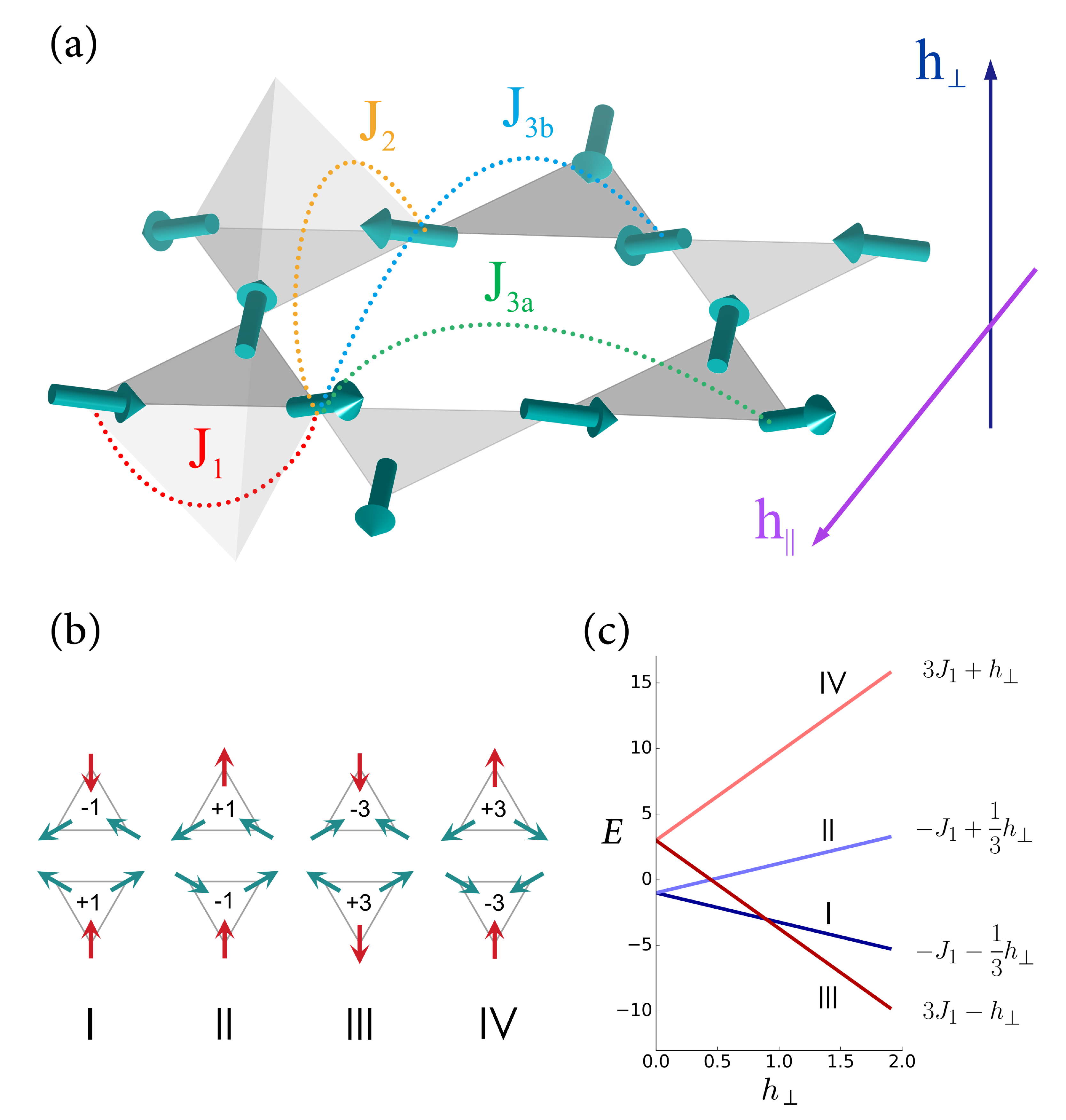}
     \caption{\label{fig:Figure1}  Canted kagome ice model. (a) The easy direction of each moment (green arrows) is pointing toward the center of a virtual tetrahedron. 
(b) Triangular configurations with possible topological charges. Red (Green) arrows represent kink (stripe) spins. 
(c) Energy-level crossing of different charges as a function of the magnitude of perpendicular field. At higher fields, charges of group I and III are more favorable than II and IV.}
\end{figure}

\section{The Model}\label{The Model}
We consider a spin model  on the two-dimensional kagome lattice  with  Ising moments canted along the local easy axes derived from the pyrochlore ice (Fig.~\ref{fig:Figure1}), 
\begin{align}
\mathcal{H}=&\bar{J}_{1}\sum_{\langle ij \rangle}\mathbf{S}_i\cdot \mathbf{S}_j + \bar{J}_{2}\sum_{\langle\langle ij \rangle\rangle}\mathbf{S}_i\cdot \mathbf{S}_j + \bar{J}_{3a}\sum_{\textrm{3rd(a)}}\mathbf{S}_i\cdot \mathbf{S}_j  
\nonumber\\ 
& + \bar{J}_{3b}\sum_{\textrm{3rd(b)}}\mathbf{S}_i\cdot \mathbf{S}_j  -  \bar{\mathbf{h}}\cdot \sum_{i}\mathbf{S}_{i}.
\end{align}
Each magnetic moment can be written as $\mathbf{S}_{i} = \mu\sigma_{i}\mathbf{d}_{\kappa (i)}$ where  $\sigma_{i} = \pm 1$ is the Ising variable and $\mathbf{d}_{\kappa(i)}$ is the unit vector along local easy axis of the $i$th spin belonging to sublattice $\kappa$.
Writing the Hamiltoinian in terms of the Ising variables $\sigma_i$, 
\begin{align} \label{eq:Hamiltonian}
          \mathcal{H} = & J_{1}\sum_{\langle ij \rangle}\sigma_{i}\sigma_{j} + J_{2}\sum_{\langle\langle ij \rangle\rangle}\sigma_{i}\sigma_{j} + J_{3a}\sum_{\textrm{3rd(a)}}\sigma_{i}\sigma_{j} \nonumber\\ 
          & + J_{3b}\sum_{\textrm{3rd(b)}}\sigma_{i}\sigma_{j} - \mathbf{h}\cdot \sum_{i}\mathbf{d}_{\kappa(i)}\sigma_{i}.
\end{align}
In the following, we use the effective exchange interactions $J_1, J_2, J_{3a}$, and $J_{3b}$ to characterize the phases, and the external magnetic field $\mathbf{h}$ is decomposed into out-of-plane and in-plane components  $\mathbf{h} = h_{\perp}\mathbf{\hat{n}} + h_{\parallel}\mathbf{\hat{k}}$ \cite{axes_detail} as depicted in Fig.~\ref{fig:Figure1}(a). 
Explicitly, the products of unit vectors and local easy axes read
\begin{equation} \label{eq:product}
     \mathbf{\hat{n}}\cdot \mathbf{d}_{\kappa} = -\frac{1}{3}
\begin{pmatrix} 1 \\ 1 \\ 1 \end{pmatrix},~
     \mathbf{\hat{k}}\cdot \mathbf{d}_{\kappa} = -\frac{\sqrt{2}}{3}
\begin{pmatrix} 1 \\ 1 \\ -2 \end{pmatrix}.
\end{equation}
In the following, we set $J_1=1$. 

\section{Ground States}\label{Ground States}
With only the antiferromagnetic nearest-neighbor interaction $J_{1} > 0$, the system has macroscopic degeneracy in the ground state. 
Each configuration in the ground-state manifold satisfies the ice rule on the kagome lattice: for both up-pointing and down-pointing triangles, the three spins must be either two-in-one-out or one-in-two-out. 
We can define the topological charge for the kagome ice as~\cite{Chern2011, Chern5718}
\begin{equation} \label{eq:charge}
     Q_{\bigtriangleup} = \sum_{i\in \bigtriangleup}\sigma_{i},\quad Q_{\bigtriangledown} = (-1)\sum_{i\in\bigtriangledown}\sigma_{i}
\end{equation}
for up-pointing and down-pointing triangles, respectively. 
The kagome-ice rule in this charge language becomes $ Q_{\bigtriangleup}, Q_{\bigtriangledown}=\pm 1$ (Fig.~\ref{fig:Figure1}(b)). 
Therefore, the nearest-neighbor Hamiltonian contains only self-energy of topological charges and describes a gas of single charges. 
The total charge density of the system is calculated through \begin{equation} \label{eq:charge_density}
     Q_{t} = \frac{1}{L^{2}}\left | \sum_{k}Q_{\bigtriangleup, k}\right |,
\end{equation}
such that a single- (triple-) charge ordered state gives $Q_{t} = +1$ $(+3)$. 
$L$ denotes the number of unit cells in each direction and the total number of spins $N = 3\times L \times L$.

\begin{figure}
     \includegraphics[width=1.0\columnwidth]{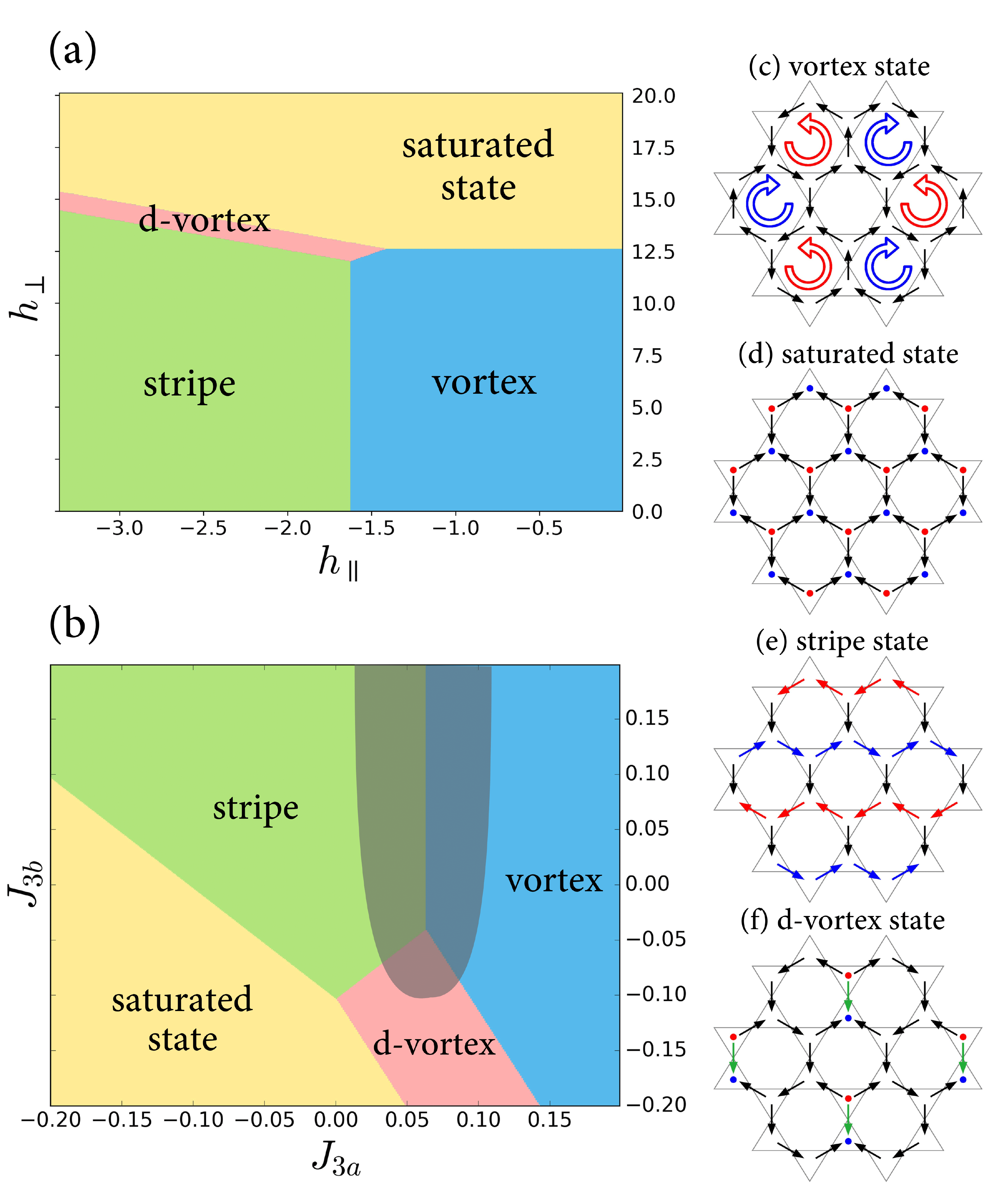}
     \caption{\label{fig:Figure2}  Energetic phase diagrams and corresponding ground-state configurations.
The phase boundaries are calculated based on the energy differences. 
The first two parameters in the Hamiltonian are fixed as ($J_{1}$,$J_{2}$) = $(1, -1/3)$.  
(a) Different field strengths with fixed ($J_{3a}$,$J_{3b}$) = $(0.05, 0)$. 
(b) Different third-neighbor coupling constants with fixed ($h_{\perp}$, $h_{\parallel}$) = $(11.76, -1.68)$. 
The snake metastable states occur at the shaded region at finite temperatures. (c)-(f) Possible ground states. The green spins in panel (f) denote flipped kink spins in the vortex state in panel (a).}
\end{figure}

The degeneracy of single-charge configurations is  lifted by  an out-of-plane magnetic field $h_{\perp}$, which acts as staggered chemical potential of triple charges (Fig.~\ref{fig:Figure1}(c)). 
Since each triangular unit contains two majority spins and one minority spin under ice rule, the configurations with majority spins aligned to the field become energetically more favorable in large perpendicular field. 
For $h_{\perp}$ larger than the critical value, the minority spin also tends to be aligned with it, leading to a $\textit{saturated state}$ of staggered triple charges (Fig.~\ref{fig:Figure2}(d)). 
These triple charges correspond to the magnetic monopoles and anti-monopoles in the pyrochlore spin ice in the [111] external field, and the saturation of charges represents the second plateau of the two-stage magnetization process described in Sec.~\ref{Introduction}.
 Thus, the canted moment in our model  allows us to imitate the magnetic field responses of the three-dimensional spin ice.
 
The projected component of canted spins on the kagome plane provides a way to perturb the kagome-ice manifold further through a small in-plane field $h_{\parallel}$.
The direction of this in-plane field is defined to point toward one vertex of the triangle in the kagome lattice, as shown in Fig.~\ref{fig:Figure1}(a).
Thus, for high-enough $h_{\parallel}$, moments on this sublattice tend to align with $h_{\parallel}$ and the other spins are arranged into the local configurations that preserve the ice rule. 
We denote these fixed spins as $\textit{kink spins}$ and the others as $\textit{stripe spins}$, as shown by different colors in Fig.~\ref{fig:Figure1}(b). 
At low temperatures below $J_1$, the states satisfying the ice-rule are characterized by polarized kink spins pointing downwards and rows of parallel stripe spins, either pointing to the left or to the right, which can be described by a local $Z_2$ order parameter. This regime is similar to the sliding phase discussed in other frustrated magnets. 
%
As $T$ is further lowered, the second-neighbor coupling $J_{2} < 0$ breaks the sliding symmetry and gives rise to an antiferromagnetic order of the parallel spin stripes. The resultant two-fold symmetric long-range ordered state is called the $\textit{stripe state}$, as depicted in Fig.~\ref{fig:Figure2}(e). 

For weaker in-plane fields, the kink spins are no longer pinned and a different long-range ordered state with $\sqrt{3}\times\sqrt{3}$ magnetic unit cell emerges. 
This so-called $\textit{vortex state}$ can be seen as an aggregation of hexagonal vortices with opposite chirality, as shown in Fig.~\ref{fig:Figure2}(c). 
The same ground state can be found in both the equivalent short-range model with in-plane moments~\cite{Wills2002} and a long-range kagome-ice model with dipolar interactions~\cite{Chern2011, Chern5718}.

Figure~\ref{fig:Figure2}(a) and (b) shows the schematic phase diagrams of the ground states found in equilibrium Monte Carlo simulations. 
Besides the known ground states mentioned above, we found a new ground state emerging within a small domain of parameter space.
The configuration of this state is depicted as Fig.~\ref{fig:Figure2}(f), and can be seen as a vortex state with columnar defects of reversed kink spins, thus we call it the $\textit{defected vortex state}$ (d-vortex state). 
Closer inspection finds one-third of the triangles in this configuration contain triple-charge defects, and the others satisfy kagome-ice rule with staggered single charges.
To our knowledge, this is a rare case that the long-range order of single and triple charges coexist.

\begin{figure*}
     \includegraphics[width=0.9\textwidth]{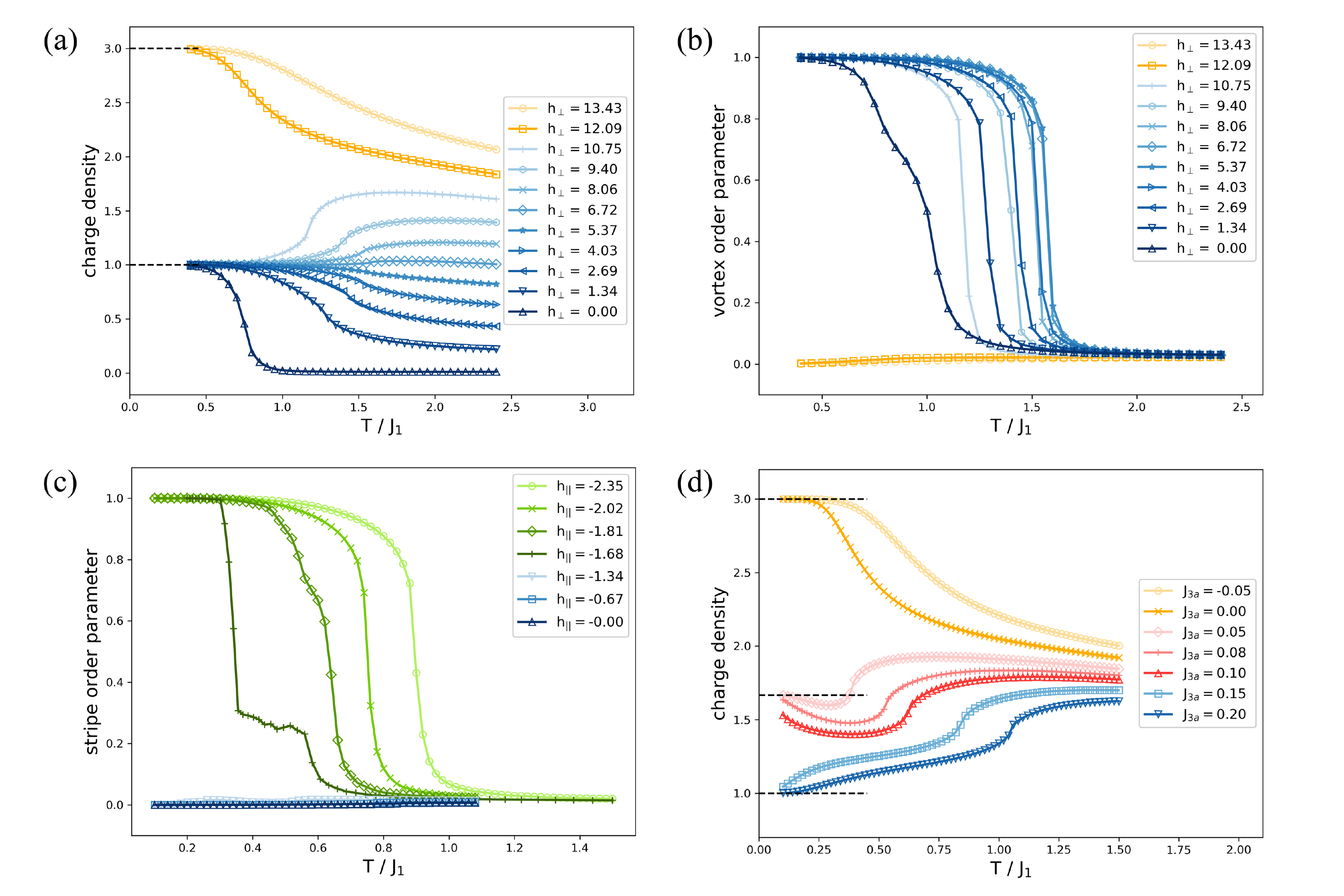}
     \caption{\label{fig:Figure3}  Equilibrium Monte Carlo results. (a) Charge density and (b) vortex-state order parameter of $J_{1}$-$J_{2}$-$h_{\perp}$ model under various $h_{\perp}$. (c) Stripe-state order parameter of $J_{1}$-$J_{2}$-$h_{\perp}$-$h_{\parallel}$ model with fixed $h_{\perp}=11.76$ and various $h_{\parallel}$ (d) Charge density of $J_{1}$-$J_{2}$-$J_{3}$-$h_{\perp}$-$h_{\parallel}$ model with fixed $J_{3b} = -0.15$ and various $J_{3a}$. The fields are chosen to be $(h_{\perp},h_{\parallel}) = (11.76, -1.68)$ as in Fig.~\ref{fig:Figure2}(b).}
\end{figure*}

The temperature and field-dependence of the charge density and various order parameters measured in the Monte Carlo simulations are presented in Fig.~\ref{fig:Figure3}. 
In the $J_{1}$-$J_{2}$-$h_{\perp}$ model, the critical field can be simply calculated from the energy difference of vortex state and saturated state as $h_{\perp, c} = 12J_{1}+12J_{3a}+6J_{3b}$ from which we obtain $h_{\perp,c} \sim 11.96$ (in the unit of $J_{1}$) for our chosen parameters, and it is consistent with the simulation results shown in Fig.~\ref{fig:Figure3}(a)-(b).
For the $J_{1}$-$J_{2}$-$h_{\perp}$-$h_{\parallel}$ model (Fig.~\ref{fig:Figure3}(c)), the critical field that drives the system into stripe state can be calculated from the energy difference of vortex state and stripe state: $h_{\parallel, c} = 3\sqrt{2}J_{2}-3\sqrt{2}J_{3a}$ and $h_{\parallel,c} \sim -1.41$. The $J_{1}$ interaction does not show up in the formula because both ground states satisfy ice rule. However, in the moderate fields, a mixed phase of vortex and stripe state is observed and numerous metastable states start to emerge and prevent the system from equilibriation. This emergent phase will be discussed extensively in the following sections.

\section{Field-quench Process\label{Relaxation process}}

Since the transitions between the various phases in Fig.~\ref{fig:Figure2}(a) and (b) are mostly first-order, mixed-phase states with complex structures often emerge at the phase boundaries especially at finite temperatures. Coupled with the underlying geometrical frustration, metastable states with unusually slow dynamics were observed particularly at the phase boundary between the stripe and the vortex states. 
%
%
Since the relaxation time scale of these metastable states is far beyond the capability of equilibrium Monte Carlo with conventional single-spin-flip or loop algorithms~\cite{melko2001}, we applied the rejection-free algorithm called the waiting time method (WTM) to investigate the slow dynamics from given initial states~\cite{dall2001faster}.
Instead of using acceptance ratio in the Metropolis algorithm, the waiting time method associates to each spin a waiting time as a local indicator of relaxation time, specifying how long a given spin has to wait before flipping.
The flipping time for each spin is given by 
\begin{equation} \label{eq:flipping_time}
     t_{i} = -\tau_{i}\log{X_{i}},
\end{equation}
where the $X_{i}$ is a random number from the uniform distribution, $X_{i} \in (0,1]$ and $\tau_{i}$ is the waiting time from the Boltzmann weight,
\begin{equation} \label{eq:waiting_time}
     \tau_{i} = \max(1, \exp{(\Delta E_{i}/T)}),
\end{equation}
where $\Delta E_i$ is the energy cost of flipping the $i^{\rm th}$ spin.
Therefore, the local flipping time $t_{i}$ is a stochastic variable from the exponential probability distribution associated with average $\tau_{i}$.
The global time $t_{g}$ is assigned to the system and updated as $t_{g} = t_{i}$ in each step after the $i^{\rm th}$ spin with lowest $t_{i}$ is flipped. 
Then, the new waiting time of the flipped spin and its affected neighbors are calculated and added to the local flipping times
\begin{equation} \label{eq:time_update}
     t^{\prime}_{i} = t_{g} - \tau^{\prime}_{i}\log{X^{\prime}_{i}},
\end{equation}
where the prime in superscript denotes that the value is generated after flipping the target spin.
To present the simulation results, the global time $t_{g}$ is used as the time scale of evolution. 
Each simulation run is executed at constant temperature and may suffer from metastability at low temperatures. 
Therefore, averaging measured observables over numerous runs is necessary for obtaining general dynamical behaviors. 

By using WTM, we study the $J_{1}$-$J_{2}$-$J_{3a}$-$h_{\perp}$-$h_{\parallel}$ model on a $L=72$ kagome lattice with 15552 spins.
The field-quench process is built up by assigning the initial configuration as saturated state that imitates the high-field condition ($h_{\perp} \gg 11.96$), and then quenching to a lower perpendicular field that favors the stripe state.
The magnitudes of magnetic fields at $t_{g}\geq 0$ are chosen to be $(h_{\perp}, h_{\parallel}) = (11.76, -1.68)$, along with third-neighbor interaction $J_{3a} = 0.05$ owing to match the energetic phase diagram in Fig.~\ref{fig:Figure2}(b).
 We found that it is enough to reach final stage of each simulation by $10^{5}$ Monte Carlo steps, and the results are the average value over 100 samples. 

\begin{figure}
     \includegraphics[width=1.0\columnwidth]{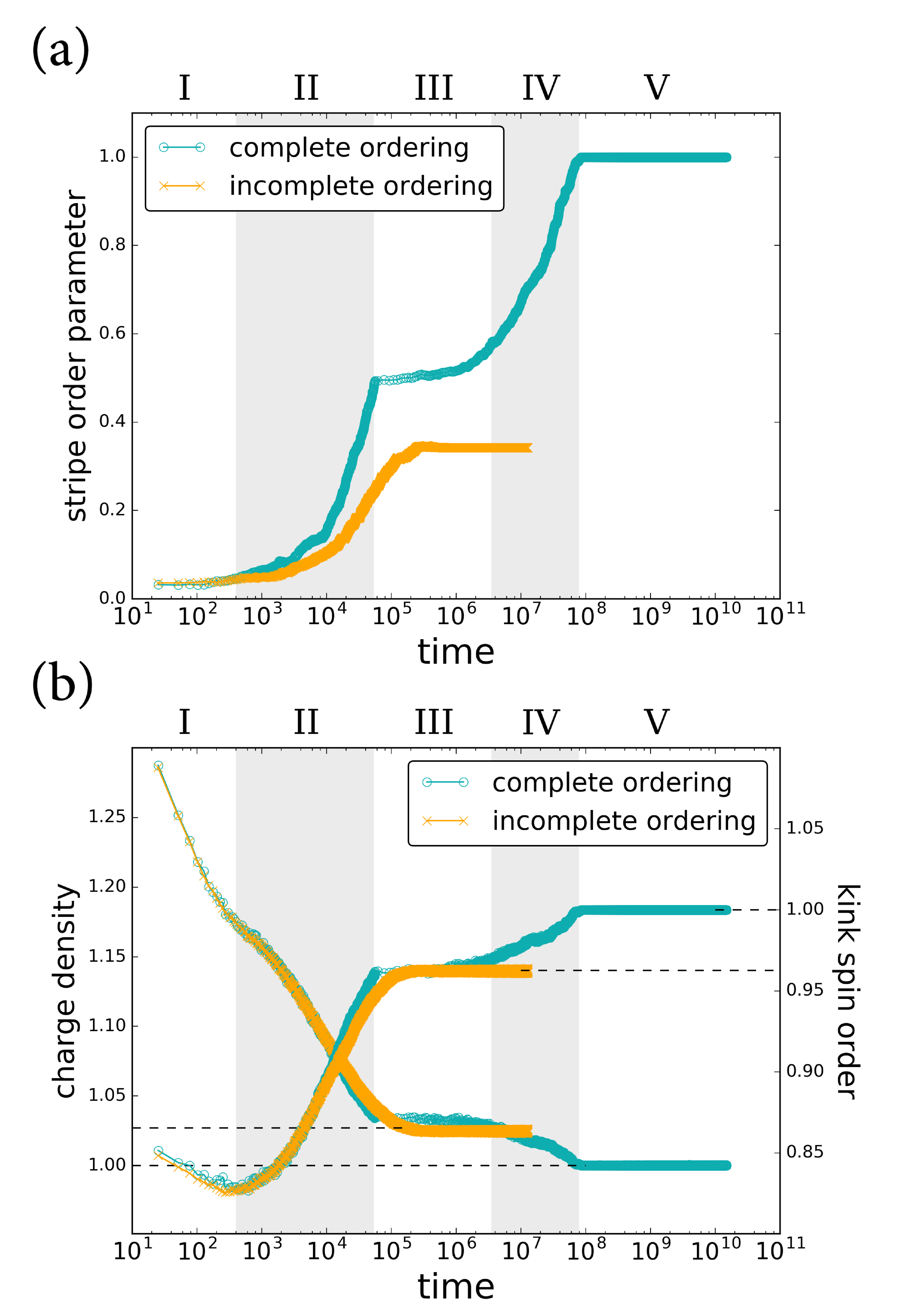}
     \caption{\label{fig:Figure4}  Field-quench process from the saturated state by waiting time Monte Carlo simulation. 
     (a) Time evolution of stripe order parameter. 
     (b) Time evolution of charge density and kink spin order parameter. 
     The green curves are averaged values from those simulations that can reach the stripe state in the end, and the orange curves are averaged values from the systems getting stuck in the snake metastable phase. 
     The time scales are measured from the global time of waiting time method.}
\end{figure}

In Fig.~\ref{fig:Figure4} we show the time evolution of the system in the field-quench scenario. 
In addition to charge density, we measured another two $\textit{ad hoc}$ order parameters to track the evolution of spin textures. 
The stripe- and kink-order parameters are defined as
\begin{equation} \label{order}
     \mathcal{O}_{s} = \left | \sum_{i = 1}^{L}\nu_{i}\left ( \sum_{j \in \nearrow, \nwarrow}\eta_{j}\sigma_{j} \right ) \right |,~\mathcal{O}_{k} = (-1)\sum_{k \in \downarrow}\sigma_{i},
\end{equation}
where $\eta = \pm 1$ for different sublattices of stripe spins, and $\nu = \pm 1$ for odd and even rows of stripes. 
The symbols $\nearrow$ and $\nwarrow$ denotes the stripe spins while $\downarrow$ represents kink spins.

As shown in the figure, the relaxation process from the saturated state can be divided into five stages.
In stage I, the charge density drops drastically by rapid pair-annihilation of triple charges.
The stripe ordering remains imperceptible since the emergent domains of single charges are randomly distributed.
In the next stage, larger striped domains are constructed by further annihilation and diffusion of triple charges.
However, in most simulation runs the system is stuck in stage III with  partial stripe ordering and shows no further evolution.
Only in very rare cases, the system in our simulation can escape from the long-lived metastable states and reach the stripe ground state in stage V. 

\section{The snake domains}\label{The snake domains}
\begin{figure*}
     \includegraphics[width=1.0\textwidth]{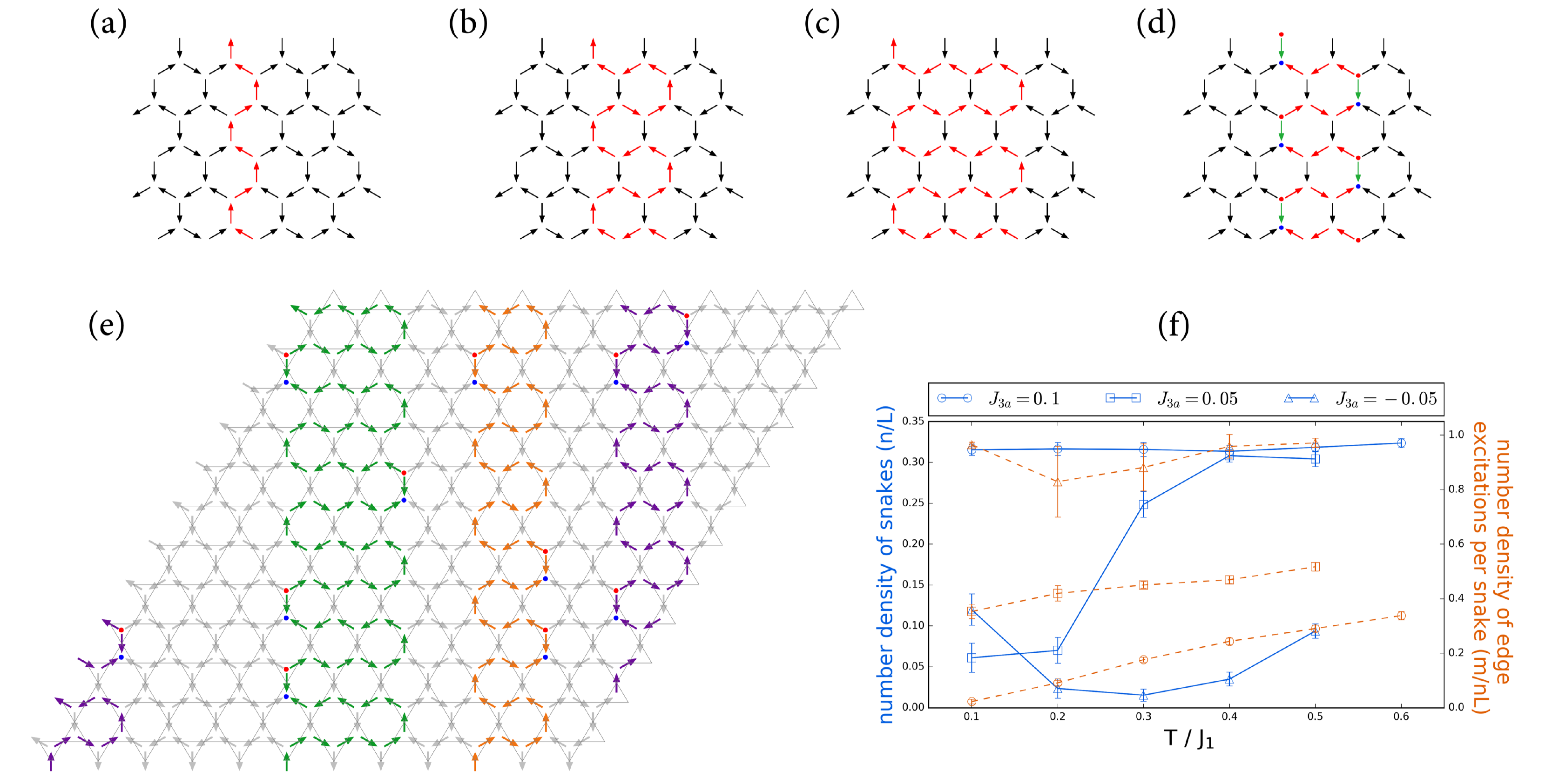}
     \caption{\label{fig:Figure5}  (a) Single string excitation on the background of the stripe state. 
     (b) Single snake of width = 3. 
     (c) Single snake of width = 5. 
     (d) Single defected snake with saturated triple charges on the edge. 
     (e) Representative configuration with multiple snake domains. 
     Red and blue dots denote the thermally driven localized triple charges. 
     Note that the magnetic domains between colored snakes are also snake domains with opposite chirality.
     (f) Temperature and $J_{3a}$ dependence of the number density of the snakes ($n/L$) and the number density of edge excitations per snake $(m/nL)$, calculated by averaging the final values of $n$ and $m$ through Eq.~(\ref{eq:nm}) from waiting time Monte Carlo simulations.}
\end{figure*}

The vortex state of the kagome ice is characterized by a broken $Z_3$ symmetry on top of the staggered $\pm 1$ charges. On the other hand, an Ising-like order parameter describes the anisotropic stripe phase. The incompatibility of these two order parameters indicates a first-order phase transition between these two ordered states. As discussed above, complex mixed-phase states often emerge at the first-order phase boundary. Here we show that the coexistence regime of the canted kagome ice exhibits an intriguing snake-like structure as well as localized ``bound'' states of monopole/anti-monopole pairs at the edges of the snake domains. The appearance of this novel snake structure is related to the emergent gauge-like sliding symmetry of the stripe phase at finite temperatures. 
%
A typical structure of a snake domain is illustrated in Fig.~\ref{fig:Figure5}(b). 
Due to the presence of further-neighbor interactions, a snake excitation is lower in energy than the string excitation (Fig.~\ref{fig:Figure5}(a)), and can be seen as a composite of single string and a series of hexagon-loop excitations.
 Since these hexagonal loops are non-winding loops and do not change the magnetization, snake and string excitation share the same topological sector and are indistinguishable by their winding numbers~\cite{macdonald2011classical, jaubert2013topological}. 
In contrast to the string excitation, the width of one snake domain can be  any odd number greater than or equal to three.
Moreover, the snakes with different widths are  degenerate in energy.
The energy change of a single snake from the background  stripe state can be written as
\begin{equation} \label{eq:snake}
     \begin{split}
          & E_{\textrm{snake}} = E_{\textrm{stripe}} + \Delta E_{s}L, \\
          & \Delta E_{s} = 12J_{2}-12J_{3a}-2\sqrt{2}h_{\parallel},
\end{split}
\end{equation}
where $L$ is the lattice dimension along one edge and $\Delta E_{s}$ denotes the energy difference per unit length.
Note that for the simulation result presented in Fig.~\ref{fig:Figure4}, the parameters are chosen to be $(J_{2}, J_{3a}) = (-1/3, 0.05)$ and $h_{\parallel} = -1.68$. 
Therefore, the Zeeman energy of in-plane field is the only term that raises the energy from the stripe state. 
Since the energy increase of forming a snake from the stripe state is proportional to the vertical dimension $L$, it is expected that a perfect snake domain is unlikely to be seen in the thermodynamic limit.
However, according to our simulation, a broken snake can only propagate through the migration of single monopoles at the end points.
Thus, the kinetic bottleneck of moving one monopole along a highly anisotropic path results in the slow relaxation of these domains even in larger system sizes.
%

At low temperatures, once a uniform snake domain is formed and wind  through the periodic boundaries, spins become very robust against thermal fluctuation except for the kink spins on the edge.
When a snake emerges from the stripe state, the ice-rule constraint forces the kink spins on the edge to align opposite to the in-plane field.
 Flipping those spins via thermal fluctuation can lower the Zeeman energy but simultaneously break the ice rule and create triple-charge pairs. 
In a large perpendicular field  as  in the WTM simulations,  cancellation of spin-spin interaction and the Zeeman energy allows the edge spins to be more easily thermally flipped, resulting in isolated triple-charge pairs. 
 Interestingly, although the edge spins are easily excited, these charges can hardly diffuse to any adjacent positions. 
 In a sense, these triple charges are not only confined but also localized, partly attributed to the stable structure of the snake domains. 
 The excitation energy of each edge spin can be shown explicitly as
\begin{equation} \label{eq:edge}
          \Delta E_{\textrm{e}}  = 8J_{1}+4J_{2}+4J_{3a}+4J_{3b}-\frac{2}{3}h_{\perp}-\frac{2\sqrt{2}}{3}h_{\parallel}.
\end{equation}
With sufficiently large $h_{\parallel}$ or $J_{3b}$, all the edge spins can be excited and aligned with the in-plane field, as shown in Fig.~\ref{fig:Figure5}(d).

For a metastable state with snake domains, there are two types of disorders.
Firstly, both the number of the snake domains and their widths can  fluctuate. 
This contributes to the destruction of the stripe order in our simulation.
Secondly, although thermal fluctuation creates localized edge excitations, the total number of these edge spins is governed by the number of snakes. 
In general, for a given configuration of $n$ snakes and $m$ edge excitations as shown in Fig.~\ref{fig:Figure5}(e), the total energy of this configuration is 
\begin{equation} \label{eq:snake_nm}
     E_{n,m} = E_{\textrm{stripe}} + n\Delta E_{s}L + m\Delta E_{e},
\end{equation}
where $n$ and $m$ are non-negative integers with the upper bounds $n\leq L/3$ and $m\leq nL$. 
Two distinctive conditions are worth considering. 
For $n = L/3$ and $m = 0$, all the snake domains have minimum width and all edge spins are not excited.
This fully compact configuration of snakes is exactly the vortex state.
In a similar way, for $n = L/3$ and $m = nL$ it is equivalent to a defected vortex state. 
Therefore, these metastable states can be regarded as mixture of the three ground states. 
Using the order parameters we measured in the simulations, we can calculate the exact values of $n$ and $m$ via
\begin{equation} \label{eq:nm}
     n = \frac{(Q_{t}-\mathcal{O}_{k})L}{2},~m = \frac{(Q_{t}-1)L^{2}}{2}.
\end{equation}
As a remark, combining the two magnetizations $M_{\perp}$ and $M_{\parallel}$ one can also reconstruct the values of $n$ and $m$, hence the $\textit{ad hoc}$ order parameters are dispensable for this task.

The temperature and $J_{3a}$ dependence of $n$ and $m$ are presented in Fig.~\ref{fig:Figure5}(f). 
In almost all the cases, the number of excited edge spins increases monotonically  as temperature rises, indicating the isolated and localized nature of these spins.
In contrast, the number density of the snakes  does not show the same trend and is changed by third-neighbor perturbation.
On account of fixed lattice size in our simulations, the number density of snake is inversely proportional to the average size of snake-like domains. 
For $J_{3a}$ = 0.1, the zero-temperature ground state is the vortex state (Fig.~\ref{fig:Figure2}(b)), and the domain structure is quite robust against thermal fluctuation, showing the saturated value in the number density of snakes (0.33) at all temperatures considered in Fig.~\ref{fig:Figure5}(f). 
However, thermal fluctuation does excite the edge spins locally, leading to the monotonic increase in the number of excited edge spins.

On the other hand, for $J_{3a} = -0.05$, larger snake domains are easier to form since stripe state is the ground state.
In this case, we notice that at intermediate temperatures thermal energy can reduce the number density of the snakes and expand the domain width, moving the system closer to the stripe state. 
Finally, when $J_{3a}$ is closer to the vortex-stripe phase boundary, the number density of the snakes shows a gradual increase with temperature, indicating that the nucleation process of the snakes is thermally enhanced. 
Therefore, although the stripe state is the zero-temperature ground state for $J_{3a} = 0.05$, at intermediate temperatures the domain structure is  actually closer to the vortex state. 
At higher temperatures, the domain structure is eventually destroyed by thermal fluctuations.

\begin{figure*}
     \includegraphics[width=1.0\textwidth]{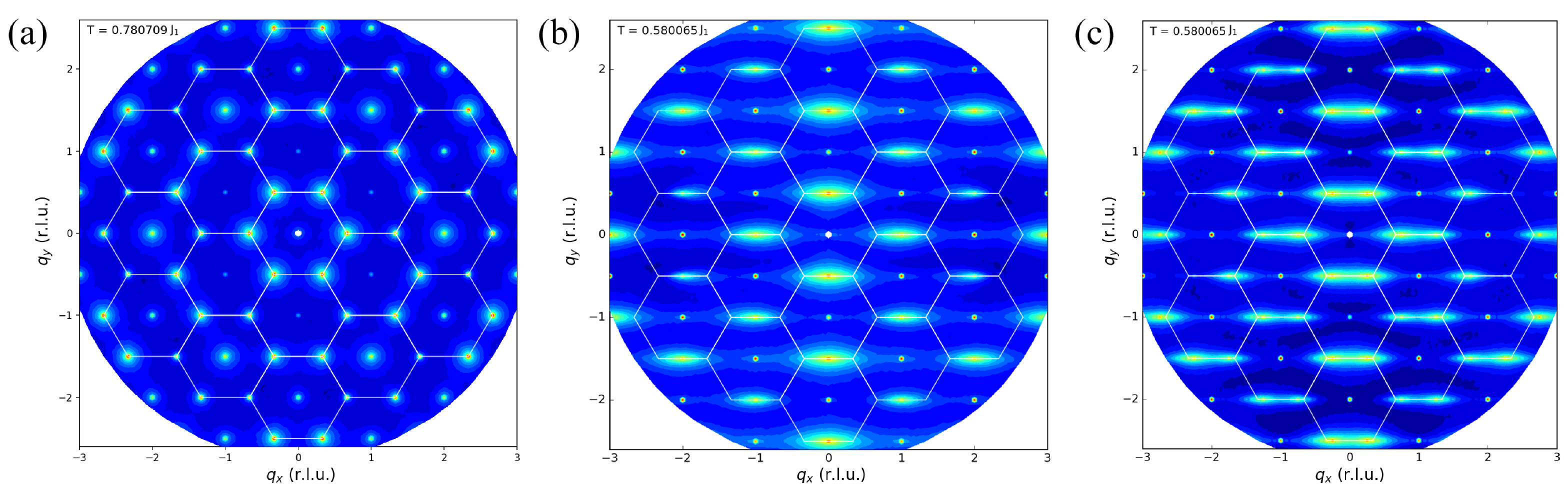}
     \caption{\label{fig:Figure6}  Magnetic structure factor of the (a) vortex state  (b) stripe state and (c) snake phase. Note that in (c), the highest-intensity points on the edge of Brillouin zone locate at somewhere between the corner and edge-center, which are the positions of Bragg peaks of vortex and stripe state, respectively.}
\end{figure*}

Finally, we remark on the possible experimental signatures of the snake phase in kagome-ice systems. 
From the  equilibrium Monte Carlo simulations, we  calculate the magnetic structure factor for different phases at finite temperatures.
On one hand, the Bragg peaks  of vortex states appear at the corners of the Brillouin zone, indicating the six-fold symmetric configurations (Fig.~\ref{fig:Figure6}(a)).
On the other hand, the presence of $h_{\parallel}$ breaks the rotational symmetry and the Bragg peaks only appear at the center of two edges with oval-shaped diffuse scattering (Fig.~\ref{fig:Figure6}(b)).
Note that this similar pattern is observed in the neutron-scattering experiment of Ho$_{2}$Ti$_{2}$O$_{7}$ single crystal under tilted [111] magnetic field~\cite{fennell2007pinch}, and later explained as the presence of long-range ordered $\mathbf{q}$ = X state, which is the three-dimensional analogy of our stripe state~\cite{Kao2016}.
However, the Bragg peaks of $\mathbf{q}$ = X state is not observed in the neutron experiment. 
For the snake phase in our model, the structure factor shows similar horizontally smeared scattering along the top and bottom edges, but the highest intensity no longer appears at the stripe position due to the presence of snake domains that breaks the  periodicity of the stripe spins (Fig.~\ref{fig:Figure6}(c)).
 In one simulation, the points with highest intensity are located at somewhere between the Bragg peaks of two long-range ordered states, indicating the frozen topological sector of kagome-ice manifold.

In the artificial spin ice systems, thermally active kagome-ice nanoarray of permalloy is successfully synthesized and studied recently. 
The equilibrium thermal behaviors such as phase transition from kagome-ice phase to long-range ordered vortex state is detected by low-energy muon spin relaxation ($\mu$SR)~\cite{Heyderman2015}, while the non-equilibrium field-quench relaxation is also investigated through time-dependent spatial correlation of moments and charges~\cite{Heyderman2017}. 
Most importantly, through the X-ray photo-emission electron microscopy and magnetic force microscopy, both the magnetic domains and local triple-charge pairs can be imaged in the artificial kagome-ice system~\cite{Heyderman2017, Zeissler2016}. 

\section{Conclusion \label{Conclusion}}
In conclusion, we study a kagome-ice model with canted spins that respond to both in-plane and out-of-plane magnetic fields. 
The inclusion of second- and third-neighbor spin-spin interactions give rise to complex texture in both spin and charge degrees of freedom. 
The various competing phases at low temperatures lead to novel snake-like structure in the coexistence regime between the vortex and stripe phases. The combination of geometrical frustration and the anisotropic nature of these snake domains give rise to robust metastable states with extremely slow relaxation dynamics after the field quench. We provide detailed structural and energetic characterizations of these snake domains in the mixed-phase states. 
%
%
In spite of its long-lived nature, spins on the edge of each domain can be easily excited by thermal fluctuation, leading to frequent creation and annihilation of triple charges.
These topological charges are localized since they can hardly overcome the energy barrier to migrate into the interior of the snake domain.
%

%
%
%

The structure of the mixed-phase states also plays an important role in the kinetics of the first-order phase transitions. For example, in the nucleation and growth scenario, the surface tension of domain walls enclosing the stable phase determines the size distribution of the nuclei as well as the nucleation speed. The elongated structure of the snake domains implies a rather anisotropic nucleation process as well as relaxation dynamics, which should be rather different from the isotropic case and is an interesting subject for future studies. 
From the experimental perspective, similar metastability and domain structure may be relevant to the absence of $\mathbf{q}$ = X Bragg peak in neutron scattering experiments~\cite{fennell2007pinch}. 
Also, the recently discovered kagome-layered compound Dy$_{3}$Mg$_{2}$Sb$_{3}$O$_{14}$~\cite{paddison2016emergent} confirms the charge-ordering transition of kagome ice and can potentially demonstrate the metastability and charge texture when external field is applied.
\begin{acknowledgements}

This work was supported  in part by the Ministry of Science and Technology (MOST) of Taiwan under Grants No. 105-2112-M-002-023-MY3,  107-2112-M-002 -016 -MY3 (WHK, YJK) ,  and 108-2918-I-002 -032 (YJK). This work was performed in part at Aspen Center for Physics, which is supported by National Science Foundation grant PHY-1607611, and was partially supported by a grant from the Simons Foundation.  
We are also grateful to the National Center for High-performance Computing
for computer time and facilities. 
\end{acknowledgements}
\bibliographystyle{apsrev4-1}
\bibliography{snakeref} 

\begin{thebibliography}{35}
\bibitem [{axe()}]{axes_detail}%
  \BibitemOpen
  \href@noop {} {}\bibinfo {note} {In order to compare the canted kagome-ice
  model with pyrochlore spin ice, it is worth noting that $h_{\perp}$ and
  $h_{\parallel}$ refer to the field strengths along the $[111]$ and
  $[\bar{1}\bar{1}2]$ crystallographic directions of pyrochlore, respectively.
  Therefore, the negative value of $h_{\parallel}$ in our model corresponds to
  the experimentally interesting negative tilt of $[111]$ field toward
  $[11\bar{2}]$ direction.}\BibitemShut {Stop}%
\end{thebibliography}
\end{document}